%% file: main.tex
\documentclass[final,3p,times]{elsarticle}

\usepackage{graphicx}
\usepackage{caption}
\usepackage{subcaption}

\usepackage{amssymb}

\usepackage{lineno}

\usepackage{amsmath}
\usepackage{amsfonts}
\usepackage{amssymb}
\usepackage{booktabs}
\usepackage{enumitem}
\usepackage{siunitx}
\usepackage{mhchem}
\usepackage{float}
\usepackage{listings}
\usepackage{hyperref}
\hypersetup{
      colorlinks=true,
      linkcolor=blue,
      filecolor=magenta,
      urlcolor=cyan,
}

\newcommand{\qe}{\textsc{Quantum ESPRESSO}}
\newcommand{\express}{\texttt{express}}
\newcommand{\Aiida}{\texttt{Aiida}}
\newcommand{\atomate}{\texttt{atomate}}
\newcommand{\Python}{\texttt{Python}}
\newcommand{\Julia}{\texttt{Julia}}
\newcommand{\ab}{\textit{ab initio}}

\newcounter{bla}

\journal{Computer Physics Communications}

\begin{document}

\begin{frontmatter}



  \title{\express{}: extensible, high-level workflows for swifter \ab{} materials modeling}


  \author[a]{Qi Zhang}
  \author[a]{Chaoxuan Gu}
  \author[b,c]{Jingyi Zhuang}
  \author[a,b,c]{Renata M. Wentzcovitch\corref{author}}

  \cortext[author] {Corresponding author.\\\textit{E-mail address:} rmw2150@columbia.edu}
  \address[a]{Department of Applied Physics and Applied Mathematics, Columbia University, New York, NY 10027, USA}
  \address[b]{Lamont--Doherty Earth Observatory, Columbia University, Palisades, NY 10964, USA}
  \address[c]{Department of Earth and Environmental Sciences, Columbia University, New York, NY 10027, USA}

  \begin{abstract}
    In this work, we introduce an open-source \Julia{} project, \express, an extensible,
    high-throughput, high-level workflow framework that aims to automate \ab{} calculations for
    the materials science community. \texttt{Express} is shipped with well-tested
    workflow templates, including structure optimization, equation of state (EOS) fitting,
    phonon spectrum (lattice dynamics) calculation, and thermodynamic property calculation
    in the framework of the quasi-harmonic approximation (QHA).
    It is designed to be highly modularized so that its components
    can be reused across various occasions, and customized workflows can be built
    on top of that. Users can also track the status of workflows in real-time, and
    rerun failed jobs thanks to the data lineage feature \express{} provides.
    Two working examples, i.e., all workflows applied to lime and akimotoite, are also
    presented in the code and this paper.
  \end{abstract}


\end{frontmatter}



{\bf PROGRAM SUMMARY}

\begin{small}
  \noindent
  {\em Program Title:} \express{}                                         \\
  {\em CPC Library link to program files:} (to be added by Technical Editor) \\
  {\em Developer's repository link:} \url{https://github.com/MineralsCloud/Express.jl} \\
  {\em Code Ocean capsule:} (to be added by Technical Editor)\\
  {\em Licensing provisions:} The GNU General Public License v3.0 (GPLv3)  \\
  {\em Programming language:} \Julia{}                                  \\
  {\em Nature of problem:} High-performance \ab{} calculation is gaining more and more popularity
  when investigating the physical and chemical properties of materials in the
  scientific community. There is a lot of \ab{} software in the market, but they are,
  more often than not, not user-friendly to new users because of their intrinsic complexity.
  Even for familiar users, dealing with daily preparation and post-analysis could be
  trivial and fallible. There are many workflow software trying to solve the problem.
  However, most of them cannot meet our expectations due to manifold reasons.
  \\
  {\em Solution method:} We developed a workflow framework that can simplify this process, i.e.,
  most of the mundane work can be replaced by writing a few lines of configurations.
  We also automated the three most-used work procedures into configurable workflows:
  equations of state fitting (structural optimization), phonon spectrum calculation, and thermodynamics calculation.
  \\

\end{small}

\input{history}
\input{design}
\input{workflows}
\input{doc}
\input{conclusions}
\clearpage
\input{tables}





\clearpage
\bibliographystyle{elsarticle-num}
\bibliography{ref.bib}







\end{document}

%% file: history.tex
\section{Introduction}\label{introduction}

Materials computations, especially of the \ab{} kind, are intrinsically complex.
For example, simulations feature many functionalities,
eventually imposing a steep learning curve on new users.
Much mundane work is performed manually, including
preparing required data,
writing and checking input,
submitting and managing calculation jobs,
and analyzing outputs.
Such tasks can be time-consuming and prone to human error.
Also, if a project is temporarily paused or handed to a new researcher,
the lack of complete tracking of previous steps can be another painful experience
for some researchers.
Luckily, there are multiple ways of mitigating challenges related to these issues.
For example, some simulation
parameters do not change during the execution cycle,
such as cell parameters,
atomic positions, and q-points.
Some other parameters have a one-to-one correspondence to each input,
e.g., pressures of interests in an equation of state calculation and fitting procedure.
Such decoupling opens opportunities for concurrent jobs in a distributed system.
These difficulties and solutions have inspired us to develop a workflow framework
to automate long and extensive \ab{} calculations.
We name this project \express, dedicated to making it direct, stable, and high-speed.

\texttt{Express} relies on previous experiences acquired in similar previous and current projects.
One of them is
the Virtual Laboratory for the Earth and Planetary Materials (VLab)
project\cite{Nacar2007,DASILVEIRA2008186}, funded by the National
Science Foundation (NSF) in 2004 at the Minnesota Supercomputing Institute.
VLab provided cyberinfrastructure for
distributed \ab{} computations through a web portal.
Users could run predefined, distributed, and interactive workflows
on a remote server via web services. Today, it still hosts plenty of online interactive workflows
\cite{mineralscloudresources}, a limited crystal structure input database
\cite{mineralsdatabase}, and a pseudopotentials/PAW datasets library\cite{pplibrary}
used in previous publications.
Unfortunately, as time went by, some of the
VLab's dependencies stopped being maintained\cite{1333397},
making it incompatible with the latest technologies.
In the meantime, we developed new and more robust versions of some VLab Java workflows.
We cherry-picked the programming techniques and public application programming interface (API)
to create a unified interface
compatible with our existing code and help reduce the workload for future development.
For example, we chose \Julia{} as our programming language,
whose object-orientation, high performance, dynamic nature, and flourishing
ecosystem create a pleasant development experience and a satisfying final product.

In this first release of \express{}, we include three workflows:
equations of state calculation, phonon spectrum calculation,
and property calculations in the framework of the quasi-harmonic approximation (QHA)\cite{Wentzcovitch2010}.
Such calculations are performed very frequently, and the methods used are well
known and tested. Therefore, these workflows can significantly reduce the need
for time-consuming human tasks, e.g., input preparation, job submission, and output analysis.

It is based on the following considerations.
The first step of most \ab{} calculations is to obtain well converged electronic structure
results and static energy-volume curve $E(V)$.
The resulting equations of state with different fitting functions are also frequently
compared to experimental results to assess the accuracy of the calculation.
Therefore, offering an EOS workflow seems a necessary first step.
Next, lattice dynamics plays a crucial role in many materials properties that
static models cannot solely explain, e.g., thermal expansion and thermal conductivity,
especially for minerals at high temperatures and pressures.
Phonon modes are also helpful in investigating
the phase stability of compounds by inspecting whether imaginary frequencies exist.
Last, when predicting thermodynamic properties of materials,
the QHA has proved computationally efficient and accurate
compared to other statistical mechanics approaches
as long as the temperatures of interest are not too high,
i.e., close to the melting temperature.
QHA results at \SI{300}{\kelvin} can also help us assess the performance of various exchange-correlation functionals,
as explained in section \ref{test-cases}.
We drew inspiration from our previously published code, including
\texttt{qha}\cite{QIN2019199} and \texttt{geothermpy}\cite{valencia2017influence},
when developing the current workflows.
The current code is well-tested, both by actual use and continuous integration systems.

Extensive efforts have already been devoted to implementing \ab{} workflows
and produced several successful libraries, including but not
limited to, \Aiida{}\cite{PIZZI2016218}, \texttt{ASE}\cite{998641},
\texttt{AFLOW}\cite{CURTAROLO2012218}, \texttt{AbiPy}\cite{VANSETTEN201839}, and
\atomate{}\cite{MATHEW2017140}. However, there is still room for
innovation. Some of this software are compatible with others or offer several
features, resulting in abstract code. On the
other hand, some packages only implement workflows for specific software.
By far, \qe{} is only supported by a few packages. Even the
largest one, \Aiida, covers limited use cases of \qe. Considering the
size of the \qe{} community, there is a great need for an advanced and eclectic workflow
ecosystem. Some packages mentioned above are implemented in \Python, a
convenient language when building a prototype project but not the most convenient one when developing a large project due to
performance issues.
We hope \express{} will be a valuable contribution to the scientific computing community, especially
those routinely performing \ab{} calculations of thermodynamic properties.

This paper is organized as follows. The next section elaborates on the design
philosophy, code composition, and basic functionalities of \express. Section
\ref{available-workflows} focuses on procedures used by the three major
workflows available here and how they facilitate typical routine
calculations. Section \ref{doc} contains the minimal technical documentation introducing
the deployment and installation of \express, input preparation, command usage,
and two working examples, lime and akimotoite. Section \ref{conclusions} offers a
closing summary.

%% file: design.tex
\section{Code design overview}\label{design}

Running calculations with external software involves many intricacies and
requires several considerations. \texttt{Express} aims
to provide a high-level interface requiring the slightest effort to learn and user-friendly
operations only. Here we briefly discuss
some design decisions
to accomplish these goals. Some design decisions may also be helpful and easily
implemented in other software, e.g., the provenance of workflows and pseudopotentials database
in \Aiida{} and \atomate. However, the programming interfaces, the
modularity of the code, and some convenient functionalities may be unique to this software.

\subsection{Three aspects of modularity to enable highly reusable
      workflows}\label{modularization}

Though it is conceivable that using \express{} as a monolithic software is
preferable by most users, we understand the need for selective installation of
its components or tweaking and adopting its functionalities into other code.
Even with that flexibility and extensibility disregarded,
writing big chunks of code is not advisable. It may result in low
readability and unexpected behaviors. Therefore, we pay great attention to
the modularity of this code during its development. The current stage of
\express{} has three almost orthogonal aspects of modularity, as described below.

\begin{figure}[H]
      \centering
      \includegraphics[height=0.55\textheight,width=\linewidth]{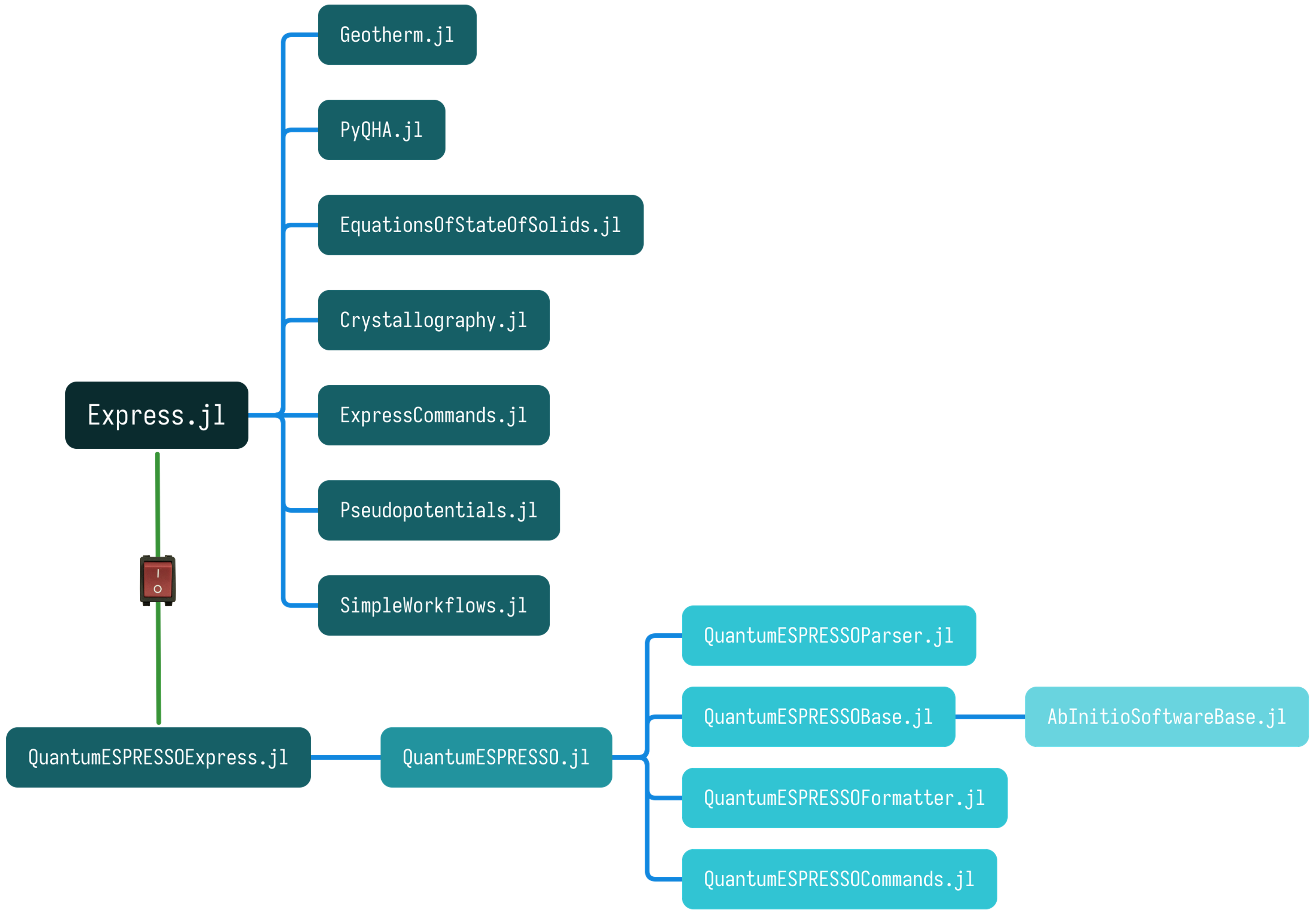}
      \caption{Main components of the \express{} project in terms of \Julia{} packages.
            The switch in this figure means that \texttt{QuantumESPRESSOExpress.jl} is a plugin
            loosely coupled to \texttt{Express.jl} and can be enabled or disabled flexibly.}
      \label{fig:components}
\end{figure}

Under the project name \express{} is a collection of \Julia{} packages,
whose core is \texttt{Express.jl}, managing and dispatching the rest.
Figurative relations of some main components of \express{} are shown in Figure \ref{fig:components},
where rectangles of the same color are at the same level, and lighter-colored rectangles
are at lower levels. High-level packages (e.g., \texttt{QuantumESPRESSO.jl})
could depend on low-level packages (e.g., \texttt{QuantumESPRESSOBase.jl}), i.e., the
high-level ones utilize the functionalities exported by the low-level ones to maximize code reuse.
Each package is released individually and has its version number to
avoid updating the whole codebase whenever there is a bug fix
or a feature enhancement. They also have separated pull request pages
for developers and skilled users to discuss and collaborate.
\Julia{}'s semantic versioning system manages their compatibility,
i.e., compatible packages are downloaded automatically,
and no human intervention is needed in most cases.
The functionalities of each package are described below.
There are several direct dependencies of \texttt{Express.jl},
which provide the basic functionalities of \express. They can be executed without
relying on each other as well.

\begin{itemize}
      \item \href{https://github.com/MineralsCloud/Express.jl}{\texttt{Express.jl}}
            provides a high-level interface to all the
            workflows, including file reading and writing, job
            creation, submission, monitoring, result retrieving, and data
            analysis. To work with specific software, install the corresponding plugin,
            e.g., \texttt{QuantumESPRESSOExpress.jl} for \qe.
      \item \href{https://github.com/MineralsCloud/ExpressCommands.jl}{\texttt{ExpressCommands.jl}}
            is a user-friendly command-line interface of
            \texttt{Express.jl} for non-developers. It installs an executable
            `\texttt{xps}' that can execute code from configuration files provided by users.
            We will review this in section \ref{running}.
      \item \href{https://github.com/MineralsCloud/EquationsOfStateOfSolids.jl}{\texttt{EquationsOfStateOfSolids.jl}}
            fits energy (or pressure) vs. volume results to equations of state,
            etc. These features are repetitively used in the equation of state
            workflow (section \ref{equation-of-state-fitting-workflow}).
      \item \href{https://github.com/MineralsCloud/Crystallography.jl}{\texttt{Crystallography.jl}}
            calculates a crystal's primitive cell (or supercell) volume from lattice parameters, finds symmetry
            operations and generates high symmetry points in the Brillouin zone, etc.
      \item \href{https://github.com/MineralsCloud/PyQHA.jl}{\texttt{PyQHA.jl}}
            is a \Julia{} wrapper of
            the \Python{} \texttt{qha} package\cite{QIN2019199}, which can calculate
            several thermodynamic properties of both single- and multi-configuration
            crystalline materials in the framework of quasi-harmonic approximation (QHA).
            The \texttt{qha} code is the foundation of the QHA workflow (section \ref{qha-workflow}).
      \item \href{https://github.com/MineralsCloud/Geotherm.jl}{\texttt{Geotherm.jl}}
            is a \Julia{} interpretation of the \texttt{Fortran} code we
            used in reference \cite{valencia2017influence}, which calculates the
            isentropic temperature/pressure gradient (geotherm) using thermodynamic properties obtained with the QHA workflow.
      \item \href{https://github.com/MineralsCloud/Pseudopotentials.jl}{\texttt{Pseudopotentials.jl}} presents
            a database for storing and querying pseudopotentials used in \ab{} calculations.
            We will elaborate on this topic in section \ref{pseudopotential}.
      \item \href{https://github.com/MineralsCloud/SimpleWorkflows.jl}{\texttt{SimpleWorkflows.jl}}
            is the skeleton of the workflow system, which
            defines building blocks, composition rules, and operation order of workflows.
            See section \ref{interfaces}.
\end{itemize}

The second modularity lies in that every workflow is software-neutral. As an example,
the \href{https://github.com/MineralsCloud/QuantumESPRESSOExpress.jl}{\texttt{QuantumESPRESSOExpress.jl}}
in Figure \ref{fig:components} is a special type of
package called ``plugin of \express{}'' for handling \ab{}
software such as \qe. Other plugins for other software are possible.
To do this, \texttt{Express.jl} defines a public protocol that
is open to extension, and the package representing the software implements
this protocol. These plugins offer a unified interface to users dealing with multiple distinct software
without modifying their code.
Though currently, only \qe{} is supported, it is straightforward to add new plugins in the future.

The dependencies of \texttt{QuantumESPRESSOExpress.jl} are listed below.

\begin{itemize}
      \item \href{https://github.com/MineralsCloud/AbInitioSoftwareBase.jl}{\texttt{AbInitioSoftwareBase.jl}}
            provides a standard API for some popular \ab{} software such as \qe.
      \item \href{https://github.com/MineralsCloud/QuantumESPRESSOBase.jl}{\texttt{QuantumESPRESSOBase.jl}}
            declares basic data types and methods
            for manipulating crystal structures, generating input files for \qe,
            error checking before running, etc.
      \item \href{https://github.com/MineralsCloud/QuantumESPRESSOParser.jl}{\texttt{QuantumESPRESSOParser.jl}}
            parses the input or output files of \qe{} to extract and analyze data.
      \item \href{https://github.com/MineralsCloud/QuantumESPRESSOFormatter.jl}{\texttt{QuantumESPRESSOFormatter.jl}}
            formats the input files of \qe.
      \item \href{https://github.com/MineralsCloud/QuantumESPRESSOCommands.jl}{\texttt{QuantumESPRESSOCommands.jl}}
            is a command-line interface that exports the commands \qe{} uses in a configurable way.
      \item \href{https://github.com/MineralsCloud/QuantumESPRESSO.jl}{\texttt{QuantumESPRESSO.jl}}
            is simply a wrapper of the types, methods, and commands defined in
            \texttt{QuantumESPRESSOBase.jl}, \texttt{QuantumESPRESSOParser.jl},
            \texttt{QuantumESPRESSOFormatter.jl},
            and \texttt{QuantumESPRESSOCommands.jl} under a common namespace.
\end{itemize}

Third, workflows should consist of atomic building blocks
that could be added, removed, or reused. It is an intuitive way that
reduces the repetition of code and human intervention.
We will elaborate on this in section \ref{database}.

\subsection{A user-friendly interface that simplifies the installation of \express{} and
      interaction with \ab{} software}\label{interfaces}

Due to the diversity of simulation software, programming languages,
and computing platforms, complete compatibility
between them is difficult or virtually impossible to achieve. If one
language or software manages to do it, it will gain the most popularity. \Python{}
is one of the candidates, but it has shown its limitations over the years,
e.g., limited performance and speed. However, an emerging language,
\Julia, was designed as an option. It has
improved interoperability with other languages, and such
versatility is continuously improving. It can now interact directly with
\Python, \texttt{C/C++}, \texttt{Java}, \texttt{R}, \texttt{MATLAB},
\texttt{Mathematica}, \texttt{Fortran}, etc., covering plenty of
practical needs. Users would spare little effort to integrate \express{} into their code if
necessary.

Apart from interoperability, the accessibility of \express{} is essential as
we understand a steep learning curve usually pushes users away.
Generally, users are held back at the moment of installing the software because of compatibility
issues. Installing
\express{} should be easy for most users, and we will discuss installation details in section
\ref{installation}.
Another way of flattening the learning curve is to provide a user-friendly yet
reusable interface.
There are four standard user interfaces: (a) directly running the \Julia{} code in read-eval-print loops (REPLs) or (b) running a
curated list of commands covering most use cases with the help of a
configuration file; (c) scripts written in one or more programming languages;
(d) the graphical user interface (GUI).
The first three interfaces are available in \express.
The first two methods require running from a local or remote terminal session.
It is possible to build a GUI, e.g., a website or an application for it in the future. But
since the majority of the code is run on remote clusters most of the time, our current focus is on
extending functionalities and improving the usability of the other interfaces.
They offer almost complete freedom when interacting with \express.

While users can benefit from exclusive \Julia{} features and libraries
not bound by a workflow-specific API,
using them requires some programming skills and familiarities with \Julia, which not everyone is expected
to have. On the other hand, the configuration file interface seems to be leveraging both
customizability and user-friendliness. Thus, it is our recommended way of using \express.
Therefore, as mentioned in section
\ref{modularization}, we ship \express{}
with an executable `\texttt{xps}' that can
handle most operations. Computational
settings are written into a human-readable file and fed to \texttt{xps} when running.
We will discuss
the details of this interface in section \ref{input-and-output-files}.
The advantages and disadvantages of each interface are listed in Table \ref{tab:interfaces}
for clarification.

\subsection{Graph model for workflow representation}\label{database}

In computer science, a workflow is often abstracted into a directed acyclic graph (DAG),
where the vertices symbolize actions and the edges specify orders of these actions.
Each vertex, doing only one thing at a time, is called an \texttt{AtomicJob} in \express.
It could be generating an input, running an executable,
reading a file, etc. \texttt{AtomicJob}s can be combined sequentially or in parallel,
forming more complex operations.
Therefore, we can make any task out of smaller existing ones.
Furthermore, any part of the workflow can be cut, branched, or looped according to need.

Apart from the flexibility, splitting a workflow into small pieces
instead of having a monolithic chunk of code has other benefits.
First, it is compliant with human perception of how
jobs are organized, making understanding, checking, and monitoring workflows
more intuitive.
It also avoids writing repetitive code, a rule of thumb in software engineering.
For example, it is often the case that we want to apply the same operation
on different input parameters, or the same operation is reused multiple times in a workflow
or in different workflows.
More importantly, it reduces the time and effort to rerun jobs because of their
loose coupling. This property is desirable
since running external software, e.g., \qe{} is usually the most computationally
expensive part.

\begin{figure}[h]  
      \centering
      \begin{subfigure}[t]{0.49\linewidth}
            \centering
            \includegraphics[width=\linewidth]{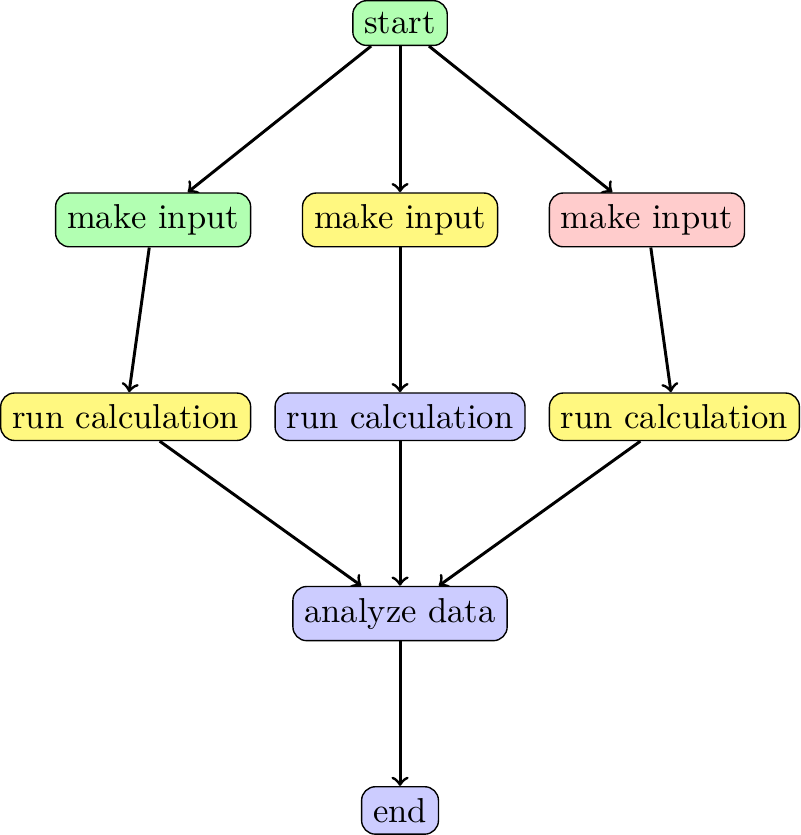}
            \caption{\label{fig:status:a}}
      \end{subfigure}
      \hfill
      \begin{subfigure}[t]{0.49\linewidth}
            \centering
            \includegraphics[width=\linewidth]{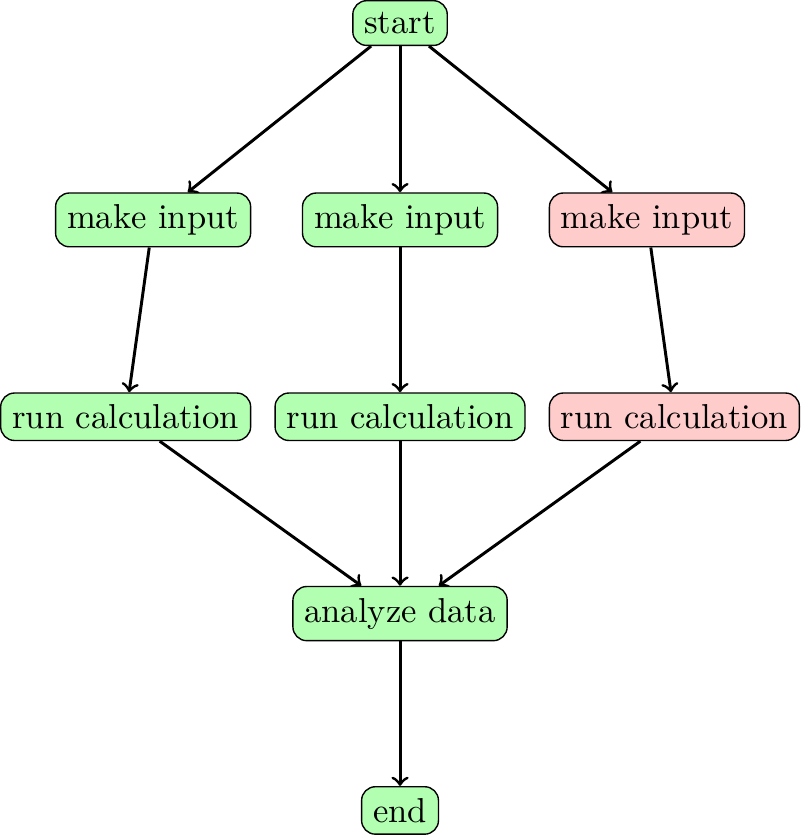}
            \caption{\label{fig:status:b}}
      \end{subfigure}
      \caption{Abstract representation of a workflow' state, plotted by \express. Two
            sequential snapshots in time, \subref{fig:status:a} and \subref{fig:status:b}, are shown.
            Figure \subref{fig:status:a} shows stage one, where some jobs are either finished,
            running, or pending. In Figure \subref{fig:status:b} (stage two), all jobs are finished.
            The final result is obtained even though one group is failed. The execution order is from
            top to bottom in both figures, and each color denotes a different status. Light green,
            pink, yellow, and light purple mean succeeded, failed, running, and pending,
            respectively.}
      \label{fig:status}
\end{figure}

Tracking job status, input parameters, and output results manually
are time-consuming and error-prone.
In \express, each job' status (whether it succeeded, failed, is running or pending),
input, and output in the workflow are tracked by an isomorphic DAG.
The DAG is saved to a file while running, making it possible to rerun the
interrupted or failed steps after restarting or fixing the reported errors. The
file can also be stored in a database for future reference or sharing information
with colleagues. Figure \ref{fig:status} shows the representation of
a simplified running workflow where each ellipse denotes a job and each
color labels a status.
In this chart, jobs are divided into three concurrent groups, where
the third one throws an error in the first step.
Therefore, its direct descendants will also fail.
Whether we will still get the final
result depends on whether the third group is decisively influential.
The next step is either to fix the error and rerun the workflow or disregard that group.
Real-world cases are usually more complex but follow the same logic.

\subsection{Pseudopotential database}\label{pseudopotential}

As mentioned in section \ref{introduction}, VLab has a limited pseudopotential library.
To follow this tradition, we integrate an optionally installable pseudopotential database
into package \texttt{Pseudopotentials.jl}.
The database lists all available pseudopotentials from PSlibrary\cite{DALCORSO2014337}
with some of their properties.
It is a local binary file in JLD2 format (an HDF5-compatible file format)
storing a tabular data structure \texttt{DataFrame}
(implemented in \texttt{DataFrames.jl}\cite{john_myles_white_2020_4282946}).
This combination guarantees it is easy to filter, aggregate, sort, insert, delete,
and perform other database operations while being very efficient.
No prior knowledge of databases is required to operate.
Some \Python{} users may find it has a similar syntax to \texttt{pandas}.
Users may also share it with colleagues.
Before running calculations, the pseudopotentials required in inputs will be
automatically downloaded from the database if the corresponding items exist.
If the files are not found either in the database or locally, the workflow being executed
will throw an exception immediately. This mechanism highlights the error and
avoids wasting computing resources since potential bugs are found before running on
\ab{} software.
We also provide a parser that could parse files of unified pseudopotential format (UPF),
which \qe, \texttt{ABINIT}, \texttt{gpaw}, etc., adopt.
Please see its official documentation\cite{Pseudopotentialsjl}
for more details.

%% file: workflows.tex
\section{Available workflows}\label{available-workflows}

\begin{figure}[H]
      \centering
      \includegraphics[width=0.8\textwidth]{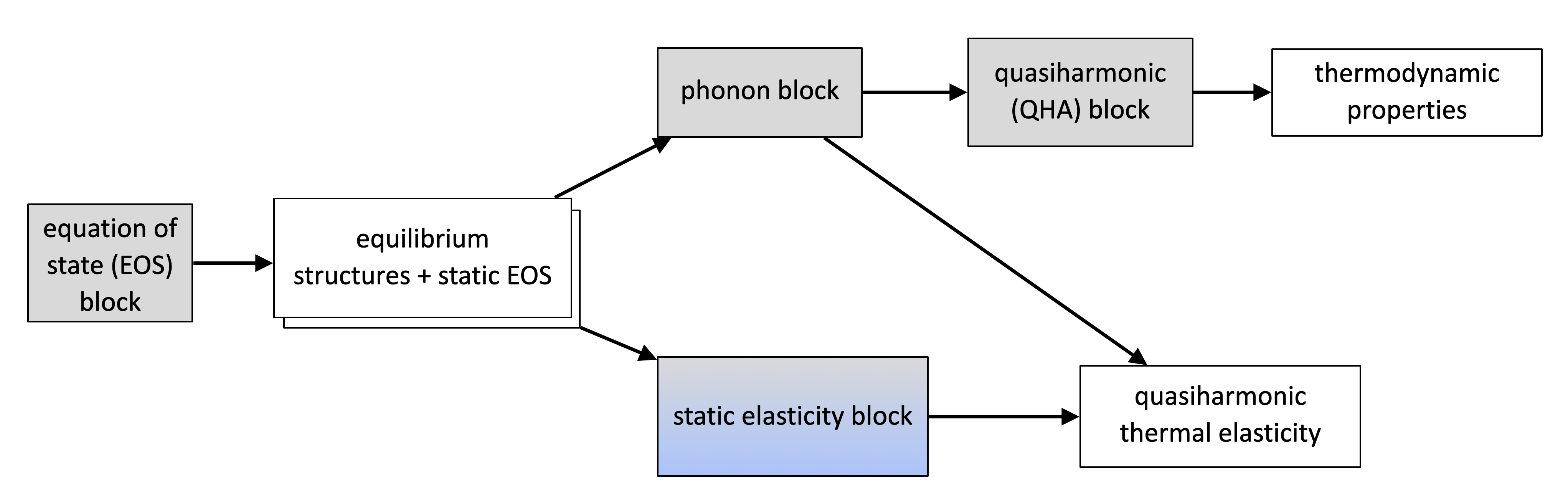}
      \caption{High-level schematic representation of the workflows.}
      \label{fig:wfls}
\end{figure}

Figure \ref{fig:wfls} presents a high-level overview of the workflows we include
in \express. Each light-gray block denotes a workflow we have and will be documented
below. The static elasticity block is not currently contained in \express{} but will be
released in the near future. A white block means the results from a previous block.
Starting from the equation of state (EOS) workflow, we get a series of equilibrium
structures. Then we could either run a phonon workflow
followed by a QHA workflow to get thermodynamic properties of a material,
or we could obtain thermoelasticity from the static elasticity workflow (whose
algorithm is proposed in reference \cite{PhysRevB.83.184115}) combined with the
QHA workflow.

As mentioned, the workflows are highly modularized, which are ready to be
separated, chained, and customized according to practical needs. For example, we could
join the EOS, phonon, and QHA workflows into one. Some workflows can have
more than one implementation, as we will explain in the following subsections.
\qe{} will be treated as the default target software unless otherwise stated.

\subsection{Equation of state (EOS) workflow}\label{equation-of-state-fitting-workflow}

\begin{figure}[H]
      \centering
      \includegraphics[width=0.8\textwidth]{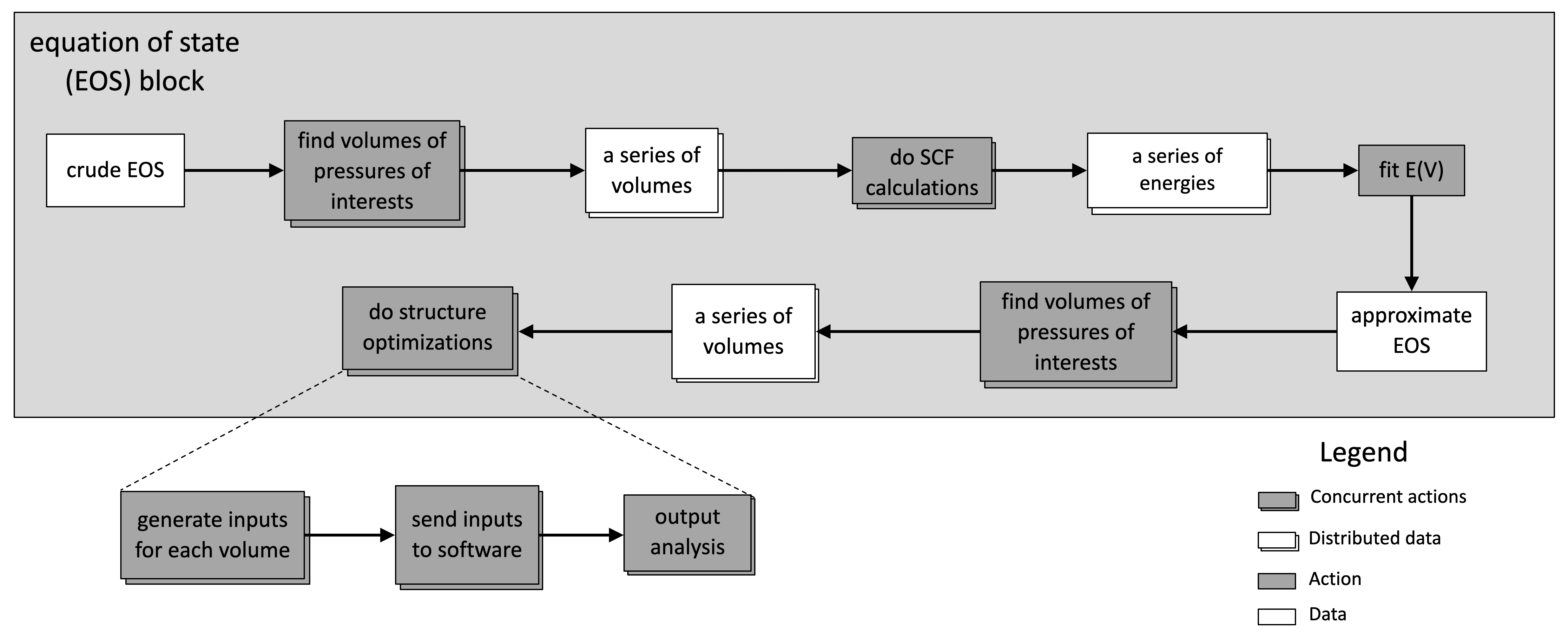}
      \caption{Visual depiction of the ``parallel'' implementation of the equation of state (EOS) workflow.}
      \label{fig:eos}
\end{figure}

Figure \ref{fig:eos} shows the EOS workflow, which is a zoomed-in view
of the corresponding block in Figure \ref{fig:wfls}, denoted by ``equation of
state (EOS) block''.
Here each gray block (action block) stands for a computational task, and each white block (data block)
corresponds to an input or output. Following this notation,
a gray stack means a group of concurrent tasks, while a white stack means a list
of inputs or outputs distributed to each task.
Some blocks are still simplified for legibility since they may contain some trivial steps,
especially when interacting with external software like \qe.
For example, the stack ``do structure optimizations'' actually
consists of three substates: input generation, running \ab{} simulation,
and output analysis (including reading optimized structures, fitting EOS, etc.).
The legend shares the same meaning in subsequent workflow graphs.
The complete procedure of this workflow is:

\begin{enumerate}
      \item Given the roughly approximate equation of state parameters
            and user-desired pressures, estimate the volumes
            corresponding to these pressures. This is done by numerically finding the solution $V$
            of a given EOS at a certain pressure: $P(V) - P_\text{desired} = 0$.
            In this process, isotropic volume expansion or compression is assumed.
            \item\label{it:2} Generate a series of inputs of self-consistent field (SCF) calculations at
            these volumes based on a template input file, and send them to the \ab{} software. As mentioned
            in section \ref{pseudopotential}, the required pseudopotentials will be
            automatically downloaded if possible during this process. We will not
            emphasize this again in the following text.
            \item\label{it:3} Gather energies and cell volumes from the outputs (using the parser in
            \texttt{QuantumESPRESSOParser.jl}) and fit them into
            a refined EOS. Multiple types of EOS can be chosen, as explained
            in Table \ref{tab:eos-config}. Unit conversion is often prone to human errors.
            Thus, we allow users to specify the units of the initial
            EOS parameters, and the fitting algorithm will return values with the same units.
      \item Save the intermediate results. Then repeat step 1, find the volumes
            corresponding to user-desired pressures using the new EOS.
      \item As in step \ref{it:2}, generate a series of inputs for structure
            optimizations at these volumes and run \ab{} calculations.
      \item As in step \ref{it:3}, fit the final EOS and get the
            equilibrium structures. Save the results to a file for future reference.
\end{enumerate}

There is also a more coupled, sequential way of sampling data points, as shown in Figure \ref{fig:eos2},
i.e., optimizing structures in increasing order of pressures where the output of the
$n$th step is used as the input of the $(n + 1)$th step.
This method may take a longer
time than the first one, given its sequential computation nature. Its complete procedure is:

\begin{figure}[H]
      \centering
      \includegraphics[width=0.8\textwidth]{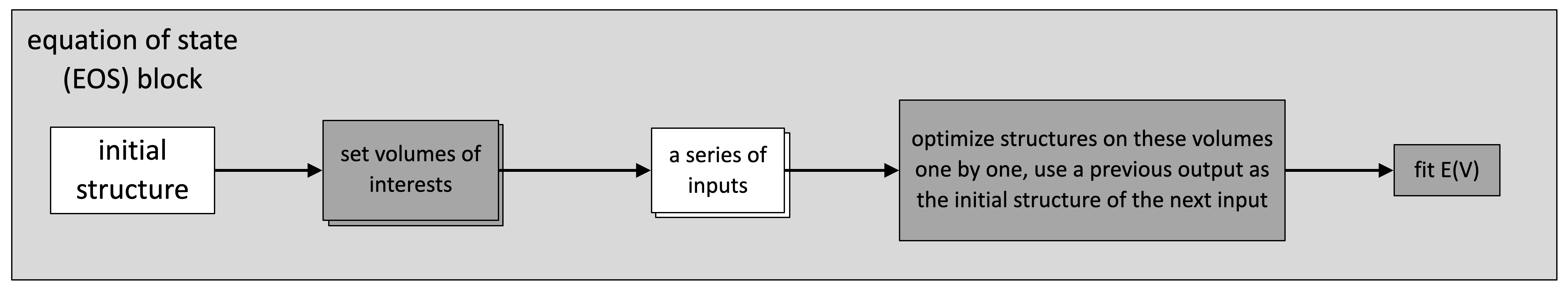}
      \caption{The ``sequential" implementation of the EOS workflow.}
      \label{fig:eos2}
\end{figure}

\begin{enumerate}
      \item Read the template file providing the initial structure and other parameters.
      \item Set desired volumes as optimization goals and generate a series of inputs with information from
            that template.
      \item Optimize the structure of the first input, use the output structure
            as the initial structure for the next input, repeat this step for the remaining
            volumes until all structures are optimized.
      \item Fit an equation of state to energy vs. volume data.
\end{enumerate}

This method is supplementary to the first one.
It is often used when a crude EOS is unknown, especially when working on a new
crystal structure where no calculations nor experiments have been performed.

\subsection{Phonon workflow}\label{phonon-workflow}

\begin{figure}[H]
      \centering
      \includegraphics[width=0.8\textwidth]{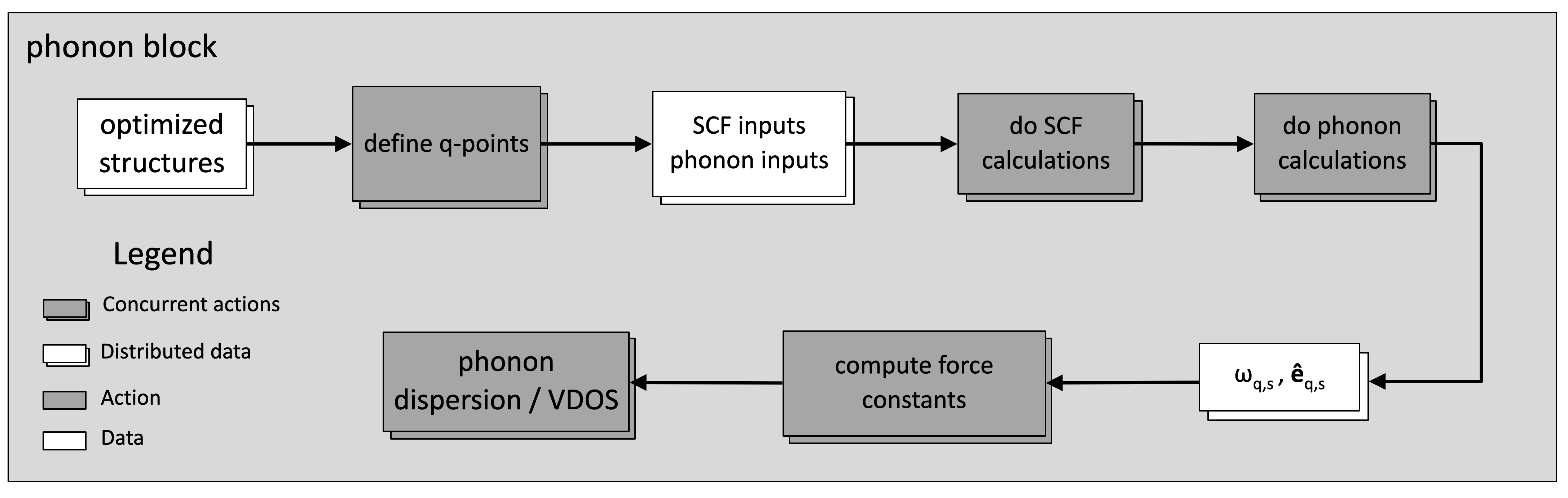}
      \caption{Graphical representation of the phonon workflow.}
      \label{fig:phonon}
\end{figure}

As shown in Figure \ref{fig:wfls}, the phonon block is indispensable whether the goal is
to obtain thermodynamic or thermoelastic properties.
Also, the effects of electronic thermal excitations on phonon frequencies
and their implications for the thermodynamic
properties are addressed in our recent publication\cite{PhysRevB.103.144102}.
The code (\texttt{pgm}) will be published and integrated into a future version of \express{}.
The procedure of the phonon workflow is:
\begin{enumerate}
      \item Start from one or more equilibrium crystal structures, which are not
            necessarily but likely to be optimized by an EOS workflow; define q-points
            automatically. Generate input files for the next step.
      \item Perform multiple distributed SCF and density functional perturbation
            theory (DFPT)\cite{dfptBaroni,dfptGonze} calculations to obtain dynamical matrices
            at those user-defined q-points. This step often requires distributing jobs over CPU cores
            since this is usually the
            most time-consuming step in the phonon workflow. \texttt{Express} distributes
            jobs by itself.
            However, it also allows manual configuration due to the complexity of distributed computing.
      \item (Call the \ab{} software to) perform a Fourier transform to derive force constant
            matrices from those dynamical matrices.
      \item Compute phonon frequencies in the entire Brillouin zone using
            these force constant matrices. Two types of calculations can be selected:
            path-mode, where the phonon dispersion is calculated along a high-symmetry
            q-path; and uniform-mode, where the vibrational density of states (VDOS) is sampled
            from results on a uniform q-point mesh. Switching between the two modes
            requires only a keyword in the configuration file.
\end{enumerate}

While the current workflow computes phonon frequencies using DFPT,
implementing the small displacement method to compute force constant matrices will be added in the near future.

\subsection{QHA workflow}\label{qha-workflow}

\begin{figure}[H]
      \centering
      \includegraphics[width=0.8\textwidth]{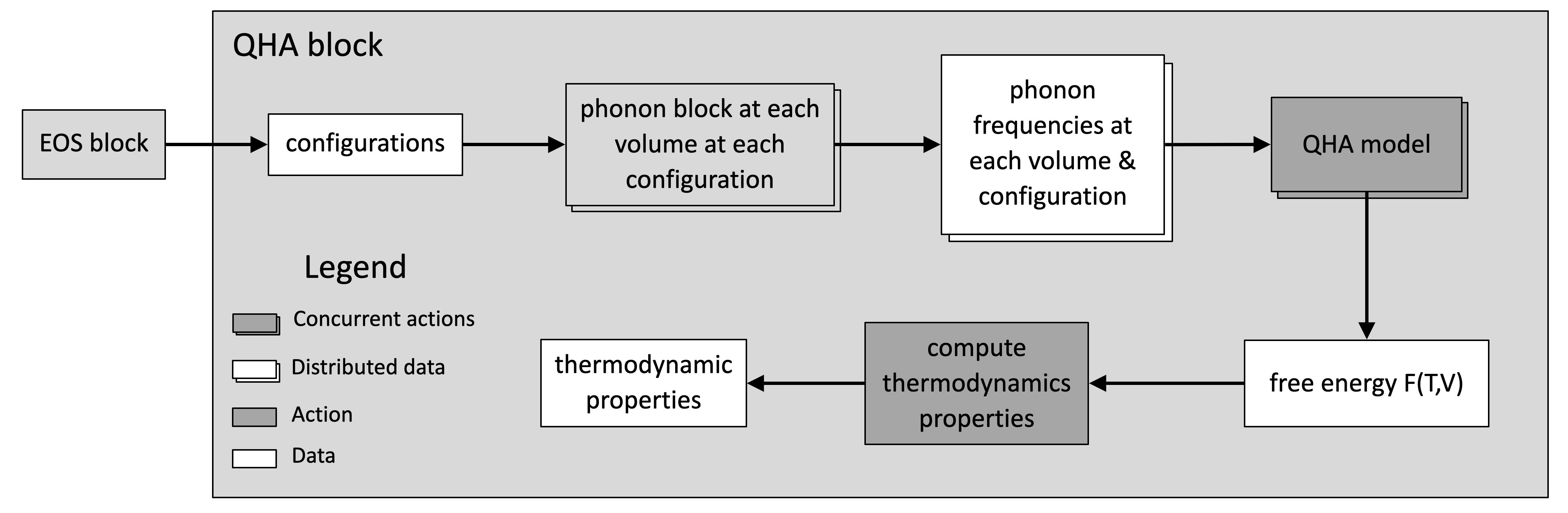}
      \caption{Schematic representation of the QHA workflow.}
      \label{fig:qha}
\end{figure}

In the \texttt{qha} package, we allow calculating the thermodynamic properties
of materials with either single or multiple configurations within a
user-specified pressure and temperature range.
The contribution of multiple configurations is not ignorable when investigating
fully or partially disordered phases or order-disorder phase boundaries, e.g., between ice-VIII
and ice-VII, high-density phases of ice\cite{UMEMOTO2010236}.
The rough steps of a QHA workflow are
\begin{enumerate}
      \item Determine whether it is a single- or multi-configuration calculation.
      \item Obtain VDOS at each statically constrained volume (of each configuration) using
            the phonon workflow. If there are imaginary frequencies, the workflow will
            warn the users that QHA might not be applicable in this case.
      \item Read input data and compute the free energy, $F(T, V)$, of
            each desired temperature $T$ and volume $V$ using the \texttt{qha} code. Interpolate $F(T, V)$ on a
            finer volume grid for better accuracy during this process.
            See equations $(2)-(8)$ in reference
            \cite{QIN2019199} for a more detailed description.
      \item Derive other thermodynamic properties, including entropy, internal
            energy, enthalpy, Gibbs free energy, thermal expansion coefficient,
            Grüneisen parameter, bulk modulus, and heat capacity, of the material from
            $F(T, V)$. Save all the results to human-readable text files, and plot them
            in PDF format if desired.
\end{enumerate}

The \texttt{qha} \Python{} code features an entry command `\texttt{qha}',
which can run with some input data and a runtime settings file.
It has been the inspiration of \express{}'s command-line interface.

%% file: doc.tex
\section{Documentations on the \express{} project}\label{doc}

\subsection{Deployment and installation}\label{installation}

Usually, the software that performs \ab{} calculations is not installed on a local
computer but on a remote server or virtual machine.
\texttt{Express} can be deployed in all these cases.
Thus, there are two situations for the deployment locations of
\express{} and the \ab{} software (for example, \qe{}): (i) they are
installed and run in the same environment, whether it is
a local computer, a remote server, or a virtual machine; (ii) they are installed in different
environments, e.g., \express{} is local and \qe{} is remote.
The latter one is a bit tricky as it requires
file transferring and calls between different systems.
By now, we recommend installing \express{} in the same environment
where the \ab{} software is installed. In general, this procedure will be smooth since
\express{} can be installed wherever \Julia{} and \Python{} can be installed,
which covers almost all operating systems and CPU architectures. So it is the assumed
method in the following text.

First, you should install \Julia. We recommend downloading it
from its official website\cite{juliadownload}. Versions higher than
\texttt{1.3}, especially \texttt{1.6}, are strongly recommended. Please follow the detailed
instructions on its website if you have to build \Julia{} from
source\cite{buildjulia}. Some
computing centers provide preinstalled \Julia. Please contact
your administrator for more information in that case.

Next, install \texttt{Express.jl} and choose one plugin for the \ab{} software.
As stated, currently, only the \texttt{QuantumESPRESSOExpress.jl} plugin is available.
Please open \Julia{}'s interactive session (REPL) and type the following lines of code:

\begin{lstlisting}[frame=single,columns=flexible]
using Pkg
Pkg.add("Express")
Pkg.add("QuantumESPRESSOExpress")
Pkg.add("ExpressCommands")
\end{lstlisting}

\noindent Then wait until it is done.

We prepared and ran many unit tests to validate the packages that constitute \express{}.
We employed several continuous integration (CI) services
(GitHub Actions, AppVeyor, Drone CI, Cirrus CI, and GitLab CI)
to run these tests on every Git commit and pull request across
multiple platforms and \Julia{} versions so that most incompatibilities and other
bugs can be identified before the official release of \express.

We host more detailed documentation on our website
(reference \cite{expressjldoc} for stable releases
and reference \cite{expressjldevdoc} for the development version) for
troubleshooting. You are also welcome to report bugs to us
\cite{expressissue}
or start a discussion if anything is unclear
\cite{expressdiscuss}.

\subsection{Input and output files}\label{input-and-output-files}

As suggested in section \ref{interfaces}, we regard the configuration file
interface as our primary way of interacting with \express.
Therefore, we will mainly describe how to prepare the input files for it here.
The usage of other interfaces,
including calling public API, is discussed in our official documentation on GitHub.
However, the same input files could be reused by other interfaces. There is no cost for switching interfaces.

We need two types of input files for each workflow: a template input for the \ab{} software and
a configuration file.
The template input' syntax may vary since it completely depends on the software,
but all workflows share the configuration file' syntax.
However, it does not mean that all
configuration files ask for the same items. For each type of workflow,
the items change a little. The complete set of allowed items
for each kind of workflow is listed in tables \ref{tab:eos-config}--\ref{tab:qha-config}.
The code block below shows a typical configuration file for an EOS workflow.
To view more real-world examples, please visit our GitHub repository
\cite{expressexamples}.
When running the workflows, \express{} will automatically generate more input
files based on the specifications in the template
input and the configuration file. This action saves users a lot of time and
labor since each input file usually varies slightly from the other.
Most of the input parameters are fixed in a series of calculations, e.g.,
DFT+U calculations with a constant U.
The copying,
pasting, and editing procedures are mundane and error-prone.
For example, only two variables change when setting
pressures of interests and corresponding volumes in an EOS workflow. It is very unproductive if one
has to edit those input files one by one.

Different workflows would have different outputs. But generally,
they are divided into two categories: raw data returned by the \ab{}
software and processed data, figures, or logs
that \express{} prints or plots directly. The first type of file is usually
the source of the second type. However, the second file type can also be the source of files
of the first type, e.g., inputs, if we chain workflows together.
For instance, the optimized structures are extracted and
treated as input structures in a phonon workflow after an EOS workflow.

\begin{lstlisting}[frame=single,columns=flexible]
recipe = 'eos'
template = 'template.in'
[cli.mpi]
np = 128
[save]
status = 'status.jls'
eos = 'eos.jls'
[trial_eos]
type = 'bm3'
values = ['300.44 bohr^3', '74.88 GPa', 4.82]
[fixed.pressures]
unit = 'GPa'
values = [-5, -2, 0, 5, 10, 15, 17, 20]
\end{lstlisting}

\subsection{Running workflows}\label{running}

This section shows how to run a configuration file
and illustrates how much work can be simplified compared to running calculations manually.
After installation, the path to \texttt{xps} (\texttt{\$HOME/.julia/bin/xps} by default)
is expected to be added to the \texttt{PATH} environment variable. If not, please
refer to our documentation.

Users can interact with the configuration file interface through a terminal session.
They can also write a script and submit it to the scheduler on a high
performance computing platform after trying out the undermentioned commands on a
small sample of inputs.
When we say ``type'' or ``run'' in the following text, we mean running commands through that
session. To carry out a computation, first, users need to prepare the input files
introduced in section \ref{input-and-output-files}. Then type

\begin{lstlisting}[frame=single,columns=flexible]
    xps run <path-to-config-file>
\end{lstlisting}

\noindent where \texttt{<path-to-config-file>} is the location of the configuration file
on the file system. In an EOS workflow, the final results include outputs returned by \qe,
a list of raw data (volume-energy pairs), and a fitted EOS.
If something goes wrong, the workflow might be terminated. Its state and the error will
be saved in a file for debugging, as mentioned in section \ref{database}.
Once the bug is fixed, run \texttt{xps run <path-to-config-file>} again, and \express{} will
retry the failed jobs.
To print either input or output data in a formatted, readable form, run

\begin{lstlisting}[frame=single,columns=flexible]
    xps print <file-name>
\end{lstlisting}

\noindent where the allowed extensions of \texttt{<file-name>} are \texttt{.jls},
\texttt{.json}, \texttt{.yaml} or \texttt{.yml}, and \texttt{.toml}.
The last four extensions correspond to three human-readable data-serialization file
formats, i.e., JSON, YAML, and TOML, while \texttt{.jls} is a binary
serialization format only recognizable to \Julia.
\texttt{Express} can also plot some data, such as the fitted EOS applied to a certain
range of volumes along with the raw data. The corresponding command is

\begin{lstlisting}[frame=single,columns=flexible]
    xps plot <file-name>
\end{lstlisting}

\noindent where \texttt{<file-name>} refers to the EOS binary file with
extension \texttt{.jls}.
These are the three most important commands of \express.
These catchy commands cover all the functionalities we have promised,
including but not limited to
unit conversion, pseudopotential downloading, input validation and generation,
calculation monitoring, task distribution, gathering and analysis of results, error handling,
logging, and visualization. We hope they can facilitate tedious work as much as possible.

\subsection{Test cases}\label{test-cases}

\begin{figure}[h]  
    \centering
    \begin{minipage}[t]{.48\linewidth}
        \centering
        \includegraphics[width=\linewidth]{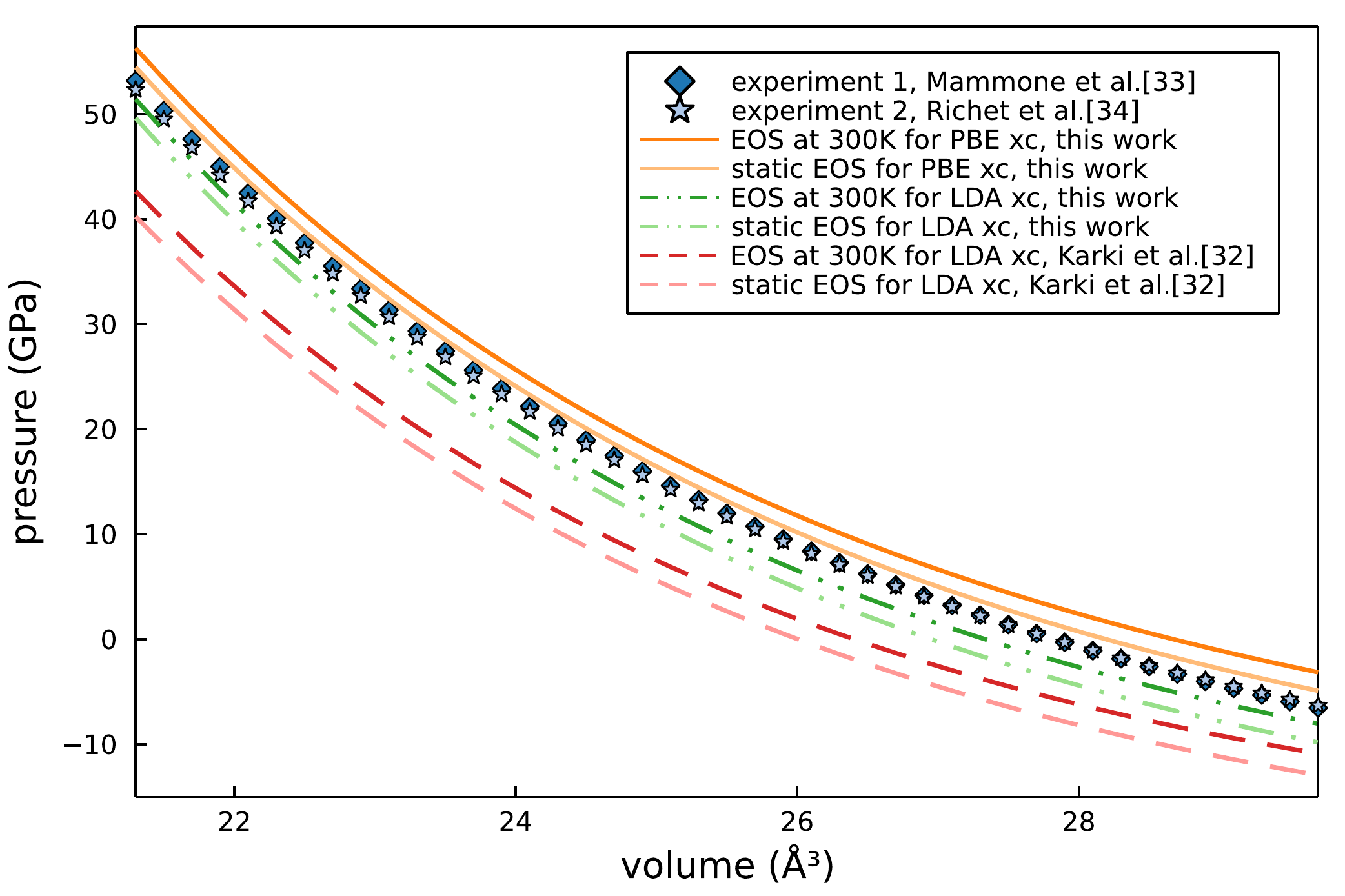}
        \caption{Compression curves for lime (\ce{CaO}). Dashed-dot-dot lines
            denote static and thermal equations of state calculated with LDA
            exchange-correlation functionals in this work, while solid lines are
            calculated with PBE-GGA exchange-correlation functionals. Dashed
            lines are from a previous \ab{} study\cite{PhysRevB.68.224304}.
            Diamonds and stars are results measured in two diamond anvil cell
            (DAC) experiments\cite{GL008i002p00140,JB093iB12p15279}.}
        \label{fig:lime}
    \end{minipage}
    \hfil
    \begin{minipage}[t]{.48\linewidth}
        \centering
        \includegraphics[width=\linewidth]{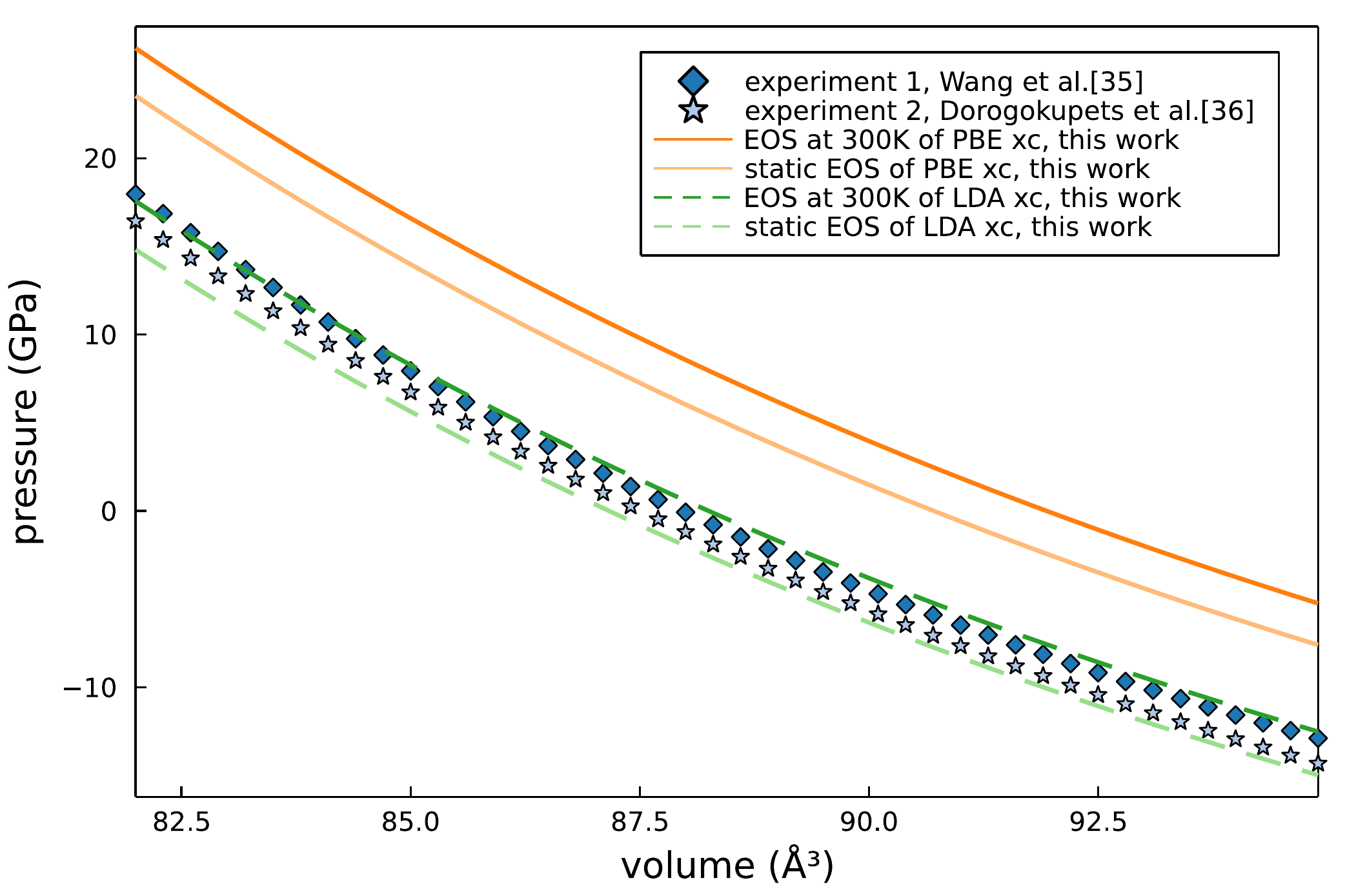}
        \caption{Compression curves for corundum-type \ce{MgSiO3}, akimotoite,
            obtained using different exchange-correlation functionals (LDA and
            PBE-GGA, dashed and solid curves) and temperatures (static and \SI{300}{\kelvin}) in this work, compared
            with experimental data (diamonds and stars) from two multi-anvil
            experiments\cite{WANG200457,DOROGOKUPETS2015172}.}
        \label{fig:ak}
    \end{minipage}
\end{figure}

These workflows are well tested, and we want to apply them to a wide range of
materials. Here we present two examples, lime, and akimotoite, with different
space groups, number of atoms per cell, atom types, etc.
We perform the EOS, phonon, and QHA workflows successively to obtain the static
and thermal properties of the two materials.
The entire procedure is wrapped in one job submitted to a compute node,
where a distribution of \qe{} is installed.
The local-density approximation (LDA) and the
Perdew--Burke--Ernzerhof generalized gradient approximation (PBE-GGA) are used
for the exchange and correlation functionals.
We did not adopt the SCAN\cite{SCAN} and the PBEsol\cite{PBEsol} functionals in these examples, but the workflows
work equally well for them as well as for DFT+U with constant U.
Figures \ref{fig:lime} and \ref{fig:ak}
show both static and thermal equations of state (third-order Birch--Murnaghan) compared with other \ab{}
calculations and experiments.
The results calculated with \express{} are in
relatively good agreement with the experimental data at room temperature,
but the LDA functional performs slightly
better than PBE.
The results are within the margin of error expected for these
functionals.

Lime (\ce{CaO}) has the rock-salt (\ce{NaCl}) structure.
Our LDA calculations (dashed-dot-dot lines) use Vanderbilt ultrasoft
pseudopotentials, while PBE calculations (solid lines) use the projector augmented wave (PAW) potentials.
We adopt cutoff energies of \SI{90}{Ry} for LDA and \SI{120}{Ry} for PBE,
and a $4\times 4\times 4$ Monkhorst--Pack k-point mesh for both cases.
Each initial structure is optimized at $8$ (\SIrange{-10}{50}{\giga\pascal}) pressures before performing DFPT
calculations on a $4\times 4\times 4$ q-point mesh. Phonon frequencies are then interpolated
on a $30\times 30 \times 30$ uniform q-mesh in the Brillouin zone.
Karki et al.\cite{PhysRevB.68.224304} performed LDA simulations
with Troullier--Martins pseudopotentials (dashed lines). They used \SI{90}{Ry} as cutoff energy,
integrated the Brillouin zone over $6$ special k-points, performed DFPT calculations on
a $4\times 4\times 4$ q-point mesh.
The diamonds and stars show two equations of state of \ce{CaO} determined by
diamond anvil cell (DAC) experiments\cite{GL008i002p00140}.

Akimotoite, or corundum-type (\ce{MgSiO3}), has trigonal symmetry (R$\bar{3}$).
As in the previous example, we carry out the LDA computations with Vanderbilt ultrasoft pseudopotentials
and PBE-GGA with PAW potentials.
In this case, the plane-wave cutoff energies are
\SI{170}{Ry} for LDA and \SI{120}{Ry} for PBE, and the Monkhorst--Pack k-point grids
are both $3\times 3\times 3$.
The q-point grids for DFPT and VDOS calculations are
$3\times 3\times 3$ and $30\times 30 \times 30$, respectively.
Four equations of state (solid and dashed lines) are also fitted using the aforementioned workflows for both
exchange-correlation functionals at the static condition and room temperature. We also compare them with two equations of state
from multi-anvil experiments\cite{WANG200457,DOROGOKUPETS2015172}.

%% file: conclusions.tex
\section{Conclusions}\label{conclusions}

In this work, we take advantage of the experiences gained from developing VLab
and other cyberinfrastructures in the past, implement the workflows we use now,
and pave the way for implementing new functionalities and integrating more atomistic simulation software for materials modeling
into our workflows in the future.
\texttt{Express} aims to make research in computational materials science
simpler, faster, and more collaborative. It helps users in the preparation of inputs,
execution of simulations, and analysis of data. It tracks the steps users performed
and can restart interrupted or failed jobs. We consider the users' difficulty
in learning new software seriously, so we tried to make it user-friendly by providing multiple
interfaces, detailed documentation, and examples.
\texttt{Express}, an open-source program distributed under the GNU General Public License,
is under active development, and
more features will be available soon.
Please follow our project page\cite{expresshome}
for future updates.

\subsection*{Acknowledgments}

This work was supported by DOE award DE-SC0019759. It used the Extreme
Science and Engineering Discovery Environment (XSEDE\cite{6866038}), which is supported by
the National Science Foundation grant number ACI-1548562 and allocation
TG-DMR180081.
The authors also thank Jiayang Wang, Michel Marcondes, Hongjin Wang, and Pedro da Silveira
for their contributions and suggestions to this code.

%% file: tables.tex
\begin{table}
    \centering
    \small
    \setlength\tabcolsep{3pt}
    \caption{Comparisons between different possible interfaces of \express.}
    \begin{tabular}{p{0.20\linewidth}p{0.10\linewidth}p{0.35\linewidth}p{0.35\linewidth}}
        \toprule
        interface                               & implemented & advantages                                                                                            & disadvantages                                               \\
        \midrule
        REPL                                    & true        & Highly customizable (It can run any code.)                                                            & \begin{itemize}
                                                                                                                                                                            \item It needs to be run in interactive mode.
                                                                                                                                                                            \item Commands are typed one by one, suitable for debugging.
                                                                                                                                                                            \item It requires some programming skills.
                                                                                                                                                                        \end{itemize} \\
        scripts                                 & true        & The same as the REPL interface, except that it can run in non-interactive mode with multiple commands & It requires some programming skills to use.                 \\
        command-line (with configuration files) & true        & \begin{itemize}
                                                                    \item More user-friendly than the previous two interfaces
                                                                    \item Highly reusable
                                                                \end{itemize}                                             & Less customizable than the previous two                                                                 \\
        GUI (web or application)                & false       & The most user-friendly interface                                                                      & Less reusable than the above three                          \\
        \bottomrule
    \end{tabular}
    \label{tab:interfaces}
\end{table}

\begin{table}
    \centering
    \small
    \setlength\tabcolsep{3pt}
    \caption{Recognizable parameters in the configuration file of
        an EOS workflow in TOML syntax.}
    \begin{tabular}{p{0.2\linewidth}p{0.8\linewidth}}
        \toprule
        Keys                                                                         & Values (with a default value if applicable)                                                                                                                                                                                                                                                                                                                                                                                                     \\
        \midrule
        \texttt{recipe}                                                              & A string that represents the type of the workflow. Allowed value is \texttt{eos}.                                                                                                                                                                                                                                                                                                                                                               \\
        \texttt{template}                                                            & The path to a template input file for a specific software. It should be on the same file system where \express{} is deployed.                                                                                                                                                                                                                                                                                                                   \\
        \midrule
        \texttt{trial\_eos}                                                          & The trial EOS contains initial values for input files generation and EOS fitting.                                                                                                                                                                                                                                                                                                                                                               \\
        \hspace{1em}\texttt{type}                                                    & A string that represents the type of the EOS. Allowed values are \texttt{murnaghan} (Murnaghan), \texttt{bm2} (Birch--Murnaghan second order), \texttt{bm3}, \texttt{bm4}, \texttt{vinet} (Vinet), \texttt{pt2} (Poirier--Tarantola second order), \texttt{pt3}, and \texttt{pt4}.                                                                                                                                                              \\
        \hspace{1em}\texttt{values}                                                  & A vector of strings that specifies each value of the EOS. The default order is $V_0$, $B_0$, $B'_0$(, $B''_0$, etc.). Units must be provided.                                                                                                                                                                                                                                                                                                   \\
        \midrule
        \parbox[t]{3em}{\texttt{fixed.pressures} or\linebreak\texttt{fixed.volumes}} &                                                                                                                                                                                                                                                                                                                                                                                                                                                 \\
        \hspace{1em}\texttt{values}                                                  & Specify the pressures or volumes. It can be a vector of numbers, or a string with the syntax \texttt{"start:step:stop"} to form an arithmetic sequence where \texttt{start}, \texttt{stop}, and \texttt{step} are numbers indicating the start, the end, and the common difference of that sequence.                                                                                                                                            \\
        \hspace{1em}\texttt{unit}                                                    & The units of pressure or volume. The pressure and volume default units are \texttt{GPa} and \verb|angstrom^3|. Allowed values for volumes are \verb|nm^3|, \verb|angstrom^3|, \verb|bohr^3|, etc. Allowed values for pressures are \texttt{Pa}, \texttt{GPa}, \texttt{TPa}, \ldots, \texttt{bar}, \texttt{kbar}, \ldots, \texttt{atm}, and the combinations of \texttt{eV}, \texttt{Ry}, \texttt{hartree}, \texttt{J}, with any unit of volume. \\
        \midrule
        \texttt{files}                                                               &                                                                                                                                                                                                                                                                                                                                                                                                                                                 \\
        \hspace{1em}\texttt{dirs}                                                    & It specifies the paths of output directories.                                                                                                                                                                                                                                                                                                                                                                                                   \\
        \hspace{2em}\texttt{root}                                                    & The path of the root directory of output files.                                                                                                                                                                                                                                                                                                                                                                                                 \\
        \hspace{2em}\texttt{pattern}                                                 & A string specifying the naming convention of the output directories. Its default value is \texttt{p=}. For example, if \texttt{fixed.pressures.values} is a vector of pressures \texttt{[10, 20, 30]} which represents the relaxations are done from \SIrange{10}{30}{GPa}, then the generated inputs and outputs will be stored in directories \texttt{p=10}, \texttt{p=20} and \texttt{p=30}.                                                           \\
        \midrule
        \texttt{save}                                                                &                                                                                                                                                                                                                                                                                                                                                                                                                                                 \\
        \hspace{1em}\texttt{status}                                                  & The path to a binary file that stores the status of the workflow.                                                                                                                                                                                                                                                                                                                                                                               \\
        \hspace{1em}\texttt{eos}                                                     & The path to a binary file that stores the fitted equations of state.                                                                                                                                                                                                                                                                                                                                                                            \\
        \midrule
        \texttt{cli}                                                                 & The command-line tools settings.                                                                                                                                                                                                                                                                                                                                                                                                                \\
        \hspace{1em}\texttt{mpi}                                                     & The configurations of the MPI software.                                                                                                                                                                                                                                                                                                                                                                                                         \\
        \hspace{2em}\texttt{np}                                                      & An integer indicating the number of processors/cores/CPUs used.                                                                                                                                                                                                                                                                                                                                                                                 \\
        \bottomrule
    \end{tabular}
    \label{tab:eos-config}
\end{table}

\begin{table}
    \centering
    \small
    \setlength\tabcolsep{3pt}
    \caption{Recognizable parameters in the configuration file of a phonon workflow in TOML syntax.}
    \begin{tabular}{p{0.15\linewidth}p{0.85\linewidth}}
        \toprule
        Keys                        & Values (with a default value if applicable)                                                                                                                                       \\
        \midrule
        \texttt{recipe}             & A string that represents the type of the workflow. Allowed values are \texttt{phonon dispersion} (phonon dispersion along a q-path) and \texttt{vdos} (phonon density of states). \\
        \midrule
        \texttt{template}           &                                                                                                                                                                                   \\
        \hspace{1em}\texttt{scf}    & The path to a template input file for an SCF calculation.                                                                                                                         \\
        \hspace{1em}\texttt{dfpt}   & The path to a template input file for a DFPT calculation.                                                                                                                         \\
        \hspace{1em}\texttt{q2r}    & The path to a template input file for a Fourier transform.                                                                                                                        \\
        \hspace{1em}\texttt{disp}   & The path to a template input file for a phonon dispersion/phonon density of states calculation.                                                                                   \\
        \midrule
        \texttt{fixed}              & The same as that of an EOS workflow.                                                                                                                                              \\
        \texttt{files}              & The same as that of an EOS workflow.                                                                                                                                              \\
        \texttt{save}               &                                                                                                                                                                                   \\
        \hspace{1em}\texttt{status} & The path to a binary file that stores the status of the workflow.                                                                                                                 \\
        \midrule
        \texttt{cli}                & The command-line tools settings.                                                                                                                                                  \\
        \hspace{1em}\texttt{mpi}    & The configurations of the MPI software.                                                                                                                                           \\
        \hspace{2em}\texttt{np}     & An integer indicating the number of processors/cores/CPUs used.                                                                                                                   \\
        \bottomrule
    \end{tabular}
    \label{tab:ph-config}
\end{table}

\begin{table}
    \centering
    \small
    \setlength\tabcolsep{3pt}
    \caption{Recognizable parameters in the configuration file of a QHA workflow in TOML syntax.}
    \begin{tabular}{p{0.15\linewidth}p{0.85\linewidth}}
        \toprule
        Keys                        & Values (with a default value if applicable)                                                                                                                           \\
        \midrule
        \texttt{recipe}             & A string that represents the type of the workflow. Allowed values are \texttt{single qha} (single configuration QHA) and \texttt{multi qha} (multiconfiguration QHA). \\
        \texttt{input}              & A path to the input file of \texttt{qha}.                                                                                                                             \\
        \texttt{pressures}          & Pressures on the dense $T-P$ gird. The same as that of an EOS workflow.                                                                                               \\
        \texttt{temperatures}       & Temperatures on the $T-P$ grid.                                                                                                                                       \\
        \hspace{1em}\texttt{values} & Specify the temperatures. It can be a vector of numbers, or a string with the syntax \texttt{"start:step:stop"} to form an arithmetic sequence.                       \\
        \hspace{1em}\texttt{unit}   & The unit of temperature. Its default value is \texttt{K}.                                                                                                             \\
        \texttt{thermo}             & Specify which thermodynamic properties are supposed to be calculated. Please refer to Table 5 in reference \cite{QIN2019199}.                                         \\
        \bottomrule
    \end{tabular}
    \label{tab:qha-config}
\end{table}